# Production of $^{239}$Pu from a natural Uranium disk and "hot rock" using a neutron howitzer

Joseph Steiner*, Aaron Anderson* and Michael De Marco**

Physics Department, Buffalo State College, 1300 Elmwood Avenue,
Buffalo, New York 14222

A neutron howitzer was used to produce $^{239}$Np from the targets of natural U and a "hot rock". An intrinsic Germanium detector enabled the observations of the gamma rays in the decay of $^{239}$Np and a determination of its half life of 2.3 days. This shows that $^{239}$Pu had been produced in both targets.

## Introduction

Neutron howitzers using a Plutonium Beryllium (Pu Be) neutron source used in conjunction with an intrinsic Germanium (Ge) detector have many uses for undergraduate studies in nuclear physics. They continue to be used to produce $^{116m}$In [1,2] and $^{198}$Au [1] from $^{115}$In and $^{197}$Au by neutron irradiation. (Undergraduate students made these measurements before attempting this study discussed here in this article.) These observations are made through measurements of gamma rays produced through the decay of these isotopes made with a Ge detector or in some cases a NaI detector. But, clearly, the narrow gamma ray lines measured by the Ge intrinsic detector made many lines easily observable in the spectrum. The resolution using the Ge detector is 2.1 keV full width half maximum (FWHM) using the decay of $^{60}$Co gamma lines. The excellent resolution using Ge detectors is clearly more important in these studies than the gain in efficiency with the use of NaI detectors.

Recently, with the renewed interest in alternate energy sources and also the threat of terrorism there has been a revival of interest in nuclear reactors and weaponry. And, therefore interest in the decay of natural Uranium (U) and production of Plutonium (Pu). We discovered that we had inherited a set of Uranium disks each about the size of a quarter (masses ~ 1.5 g) with our 99.9mCi Pu Be neutron source. Most schools are unable to obtain U disks or aren't as fortunate as we are to have inherited them. Therefore, we thought it pedagogically useful to irradiate something other than U disks that could be purchased without too much difficulty. We were able to purchase a set of "hot rocks" from United Nuclear Inc. and selected a nearly rectangular shape (~ 1 inch square by ½ inch thick) rock (mass ~ 19g) which contained, perhaps, a few percent U. These target materials were irradiated in plastic bags in the distilled water filled neutron howitzer at a distance of about 2 inches from the Pu (Be)source. They were used with a set of calibrated sources, multichannel analyzer (MCA) connected to a computer, intrinsic Ge detector and spectroscopy amplifier which enabled the observations of the gamma rays in the decay of $^{239}$Np and a determination of its half life of 55 hours. These measurements show that $^{239}$Pu had been produced in both targets.

# Experimental

## A. Uranium Disk

After measuring gamma ray spectra from terrestrial or normal background sources [3] and the non irradiated U disk mentioned above using an intrinsic Ge detector we were able to identify many of the radioisotopes. The gamma ray spectrum of natural U is shown in figure 1. In the spectrum there were mainly decay products from U such as: $^{226}$Ra, $^{228}$Th and $^{208}$Tl and $^{228}$Ac. It is important to identify background radiation before neutron irradiation. This enables us to determine which isotopes have been produced by neutron irradiation. We can also see which gamma rays will cause interference and, therefore, complications in determining which radio-isotopes were produced in the neutron irradiated samples.

Most of pure Uranium is $^{238}$U with a small amounts of $^{235}$U (<1%) and $^{234}$U (<<1%). The cross section for neutron capture in $^{238}$U is a few barns, so, it was determined that on the order of one hundred decays per second of $^{239}$U would be created in the neutron irradiation process and it might be possible to observe the production of $^{239}$Pu. The decay $^{239}$U decays ($T_{1/2}$ = 23 minutes) by beta decay with no significantly intense gamma trays to follow in its decay to $^{239}$Np. However, $^{239}$Np ($T_{1/2}$ = 55 hours, 2.3days) decays by beta decay to $^{239}$Pu ($T_{1/2}$ = 2.44 x $10^4$ years) and produces gamma rays in its decay that are measurable (106, 209, 228, and 278 keV) [5].

We first irradiated the disk of natural U for a few days and then subtracted a normalized background from natural U to observe that the 106, 209, 228 and 278 keV gamma rays were present in the spectrum.

In order to confirm $^{239}$Np production we needed to measure its half-life. This meant we needed better statistics and maximum activity because of the large background radiation from U. The production of activity of short half life followed by a long half can be produced in two ways. One is to produce directly produce $^{239}$U and wait for it to decay. This is performed by the usual experimental method:

Activity produced of $^{239}$U = $\Phi \sigma N_0 (1 - e^{-0.7\ t/T})$ where T = half life of $^{239}$U = 23 minutes,

$N_0$ = number of $^{238}$U atoms,   $\Phi$ = neutron flux of about 4 x $10^4$ n/sec-cm$^2$

$\sigma$ = 2.73 barns

In our situation this does not produce much activity because of the low neutron flux and the short time to reach saturated production. The second production method is to keep producing $^{239}$U well beyond the saturation time. This is done according to:

Activity produced of $^{239}$Np = $\Phi \sigma N_0 [ 1 - \lambda c/(\lambda c - \lambda_B)[ e^{-\lambda_B t} - (\lambda_B/\lambda c)(e^{-\lambda_C t})]]$

where: $\lambda$ is the decay constant, B = $^{239}$U and C = $^{239}$Np

This second process is more conducive to producing larger activity because it produces $^{239}$Np through decay of $^{239}$U while in saturation. We determined that about 80% of maximum activity would require about a ten day irradiation. This type of production is outlined in R.D. Evan's textbook and very nicely illustrated by the deuteron production of $^{131}$Te and its decay to $^{131}$I .[4]

The 228keV gamma ray was used to determine the half life of $^{239}$Np since it had the least interference from other background gamma rays and the largest measured intensity. The determination of the activity of the 106 keV gamma ray is complicated because of its proximity to other Np and Pu X rays and the 209keV has a small intensity and interference from $^{228}$Ac while the 278keV has gamma ray interference from the decay of $^{208}$Tl. The spectrum of irradiated natural U is shown in Figure 2. The activity of the 228keV was small and the background count was high so accumulation times were made to be six hours per day. The initial net total count under the 228keV was 30,257 decays for six hours. The typical uncertainty for a measurement was about 600 decays. It included both the uncertainty in the background and the net count. The daily measuring continued for almost two half lives. The half life of the 228keV shown in figure 3 is seen to be a straight line on a natural log scale versus time. A program supplied by Dr. Eddie Brown from Lawrence Berkeley National Laboratory (LBNL) [5] was used for the half life calculation which included uncertainty in each measurement due to both the background and the net counts. The half life was found to be (53.3 +/-1.7) hours. The half lives of the 106keV, 209keV and 278keV were determined to be in the few days range consistent with the decay of $^{239}$Np.

## B. "Hot Rock"

The gamma ray spectrum of the non irradiated "hot rock" was measured to observe its natural radioactivity. It contained many of the same gamma rays as natural U disk (see figure 1) which showed that $^{238}$U was probably present in the rock. The rock was irradiated for ten days and then the spectra were measured in the same way as the U disk. The spectra of the "hot rock" before and after irradiation are similar to that of U disk except that the produced activity of $^{239}$Np is smaller. The initial net activity of the 228keV was 3418 decays in the first six hour measurement. The decay curves for the irradiated "hot rock" and U disk are seen to be nearly parallel in figure 3. The half-life of $^{239}$Np activity measured by the 228keV gamma ray was determined in the same way as the irradiated U disk. It was measured until the uncertainty in net counts was close to the background counts. The half life was found to be (46.7 +/- 10.6) hours. The lifetime of the activity of the 106keV, 209keV and 278keV gamma rays observed in this spectrum were consistent with a half life of days.

It should be mentioned that this experiment might be possible with a set of NaI detectors but certainly much more difficult because it would require different size NaI crystal detectors to restrict the detection of gamma rays and produce narrower gamma ray lines. For instance, a sample spectrum measured by us with 3" x 3" NaI of the non irradiated rock produced FWHM lines that were too broad to do such a study.

## Conclusions

This study presents a learning experience for undergraduate students detailing U decay products using gamma ray spectroscopy with an intrinsic Ge detector. It shows the multiple step production of $^{239}$Np from natural U and/or a "hot rock" using a Pu (Be) neutron howitzer by the observation of the gamma rays and the measurement of the 2.3 day half life of $^{239}$Np. And, it shows a production process for $^{239}$Pu.

* Undergraduate physics students
**Supported by the USDOE(DE-FG02-03ER46064)

1. James L. DuBard and Alok Gambhir, "Dust of the neutron howitzer to teach nuclear physics", Amer. J. Phys. 62, 255-257(1994).
2. Stephen C. Yerian, "Comment on Dust of the neutron howitzer to teach nuclear physics", Amer. J. Phys. 62, 1150-1151(1994).
3. R.R. Finck, and R.B.R. Persson, Nucl. Instru. Meth. 135, 559(1976), and Glenn Knoll, *Radiation Detection and Measurement* ( John Wiley & Sons, second edition, 1989) pg 716-717.
4. R. D. Evans, *The Atomic Nucleus*, (McGraw-Hill, 1955) pg 486-489.
5. C.M. Lederer, J. M. Hollander, and I. Perlman, *Table of Isotopes,6 $^{th}$* ( Wiley , New York, 1967), C.M. Lederer and V.S. Shirley, *Table of Isotopes*, 7$^{th}$ edition ( Wiley , New York, 1978), R.B. Firestone, *Table of Isotopes, 8$^{th}$ edition*(Wiley, New York, 1999).
6. Private communication from Dr. Eddie Brown from LBNL.

Figure 1

This spectrum shows the gamma rays measured in the decay of natural Uranium with a Ge detector. The vertical axis (log scale) represents the decays measured in a six hour measurement and the horizontal axis shows the energy scale in keV. The arrows point to the energies (106,209,228 and 278keV) of interest. The dots show the location of the relevant gamma rays. It can be seen from the spectrum show that the 228 keV gamma ray has the least gamma ray interference.

Figure 2

This spectrum shows the gamma rays produced in the neutron irradiation of natural U measured with a Ge detector. The vertical axis (log scale) shows the decays in the first six hour measurement and the horizontal axis shows energy in keV. The arrows and black dots show the energies (106,209,228 and 278keV) of interest and show that they are more intense in the neutron irradiated U disk.

Figure 3

The plot shows the natural log of the decays or counts (vertical axis) versus time in hours (horizontal axis) for the 228keV gamma ray. Each dot represents a six hour measurement using the Ge detector. The plot shows both the decay curve for natural U and the "hot rock". The lines shown are best fit lines without consideration of the uncertainties in both the background and net counts. It can be seen that they are both straight and nearly parallel. The calculation of the best fit line shown here is consistent with the computer calculation of the half life of 55 hours for both lines as detailed in the text.

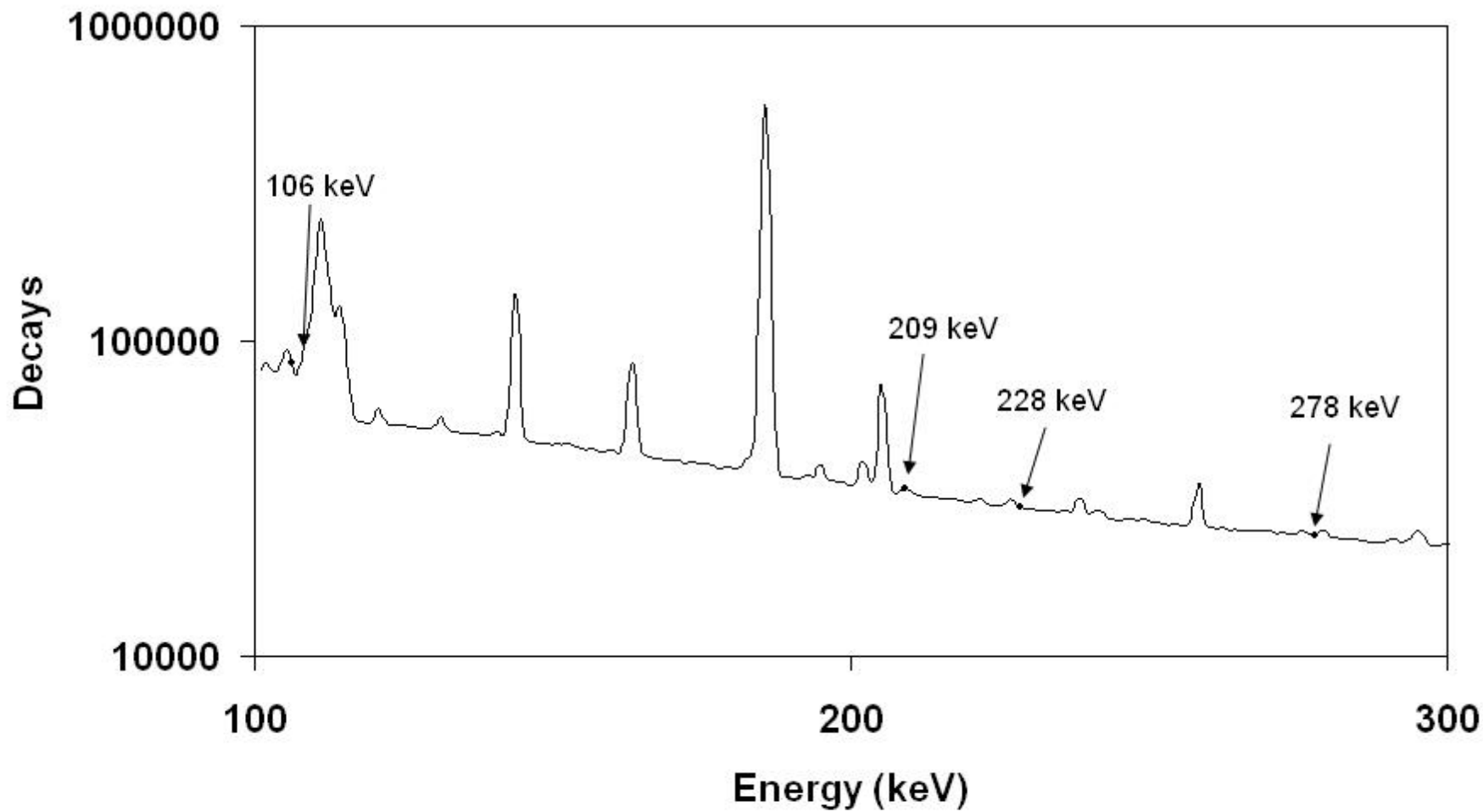

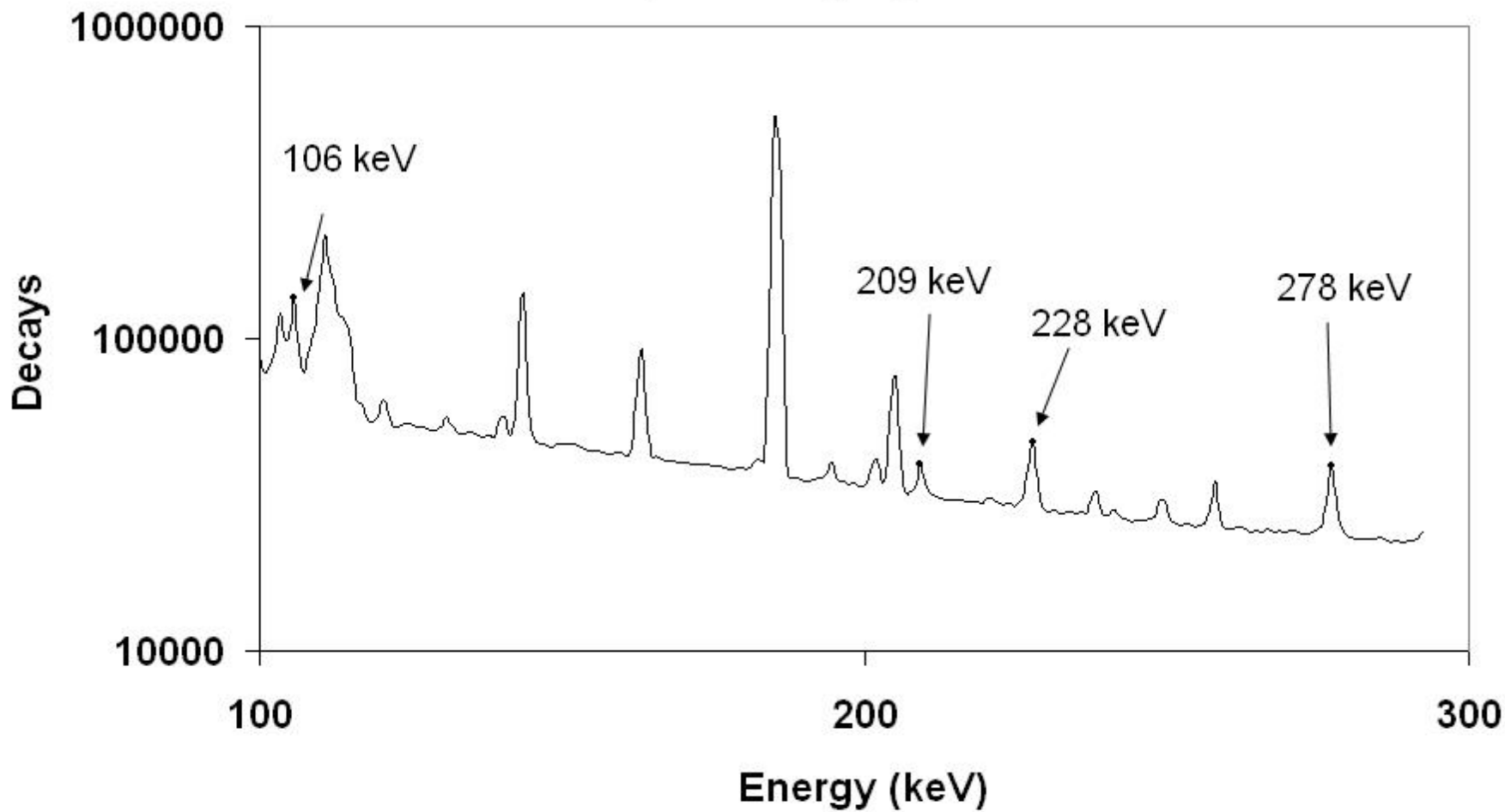

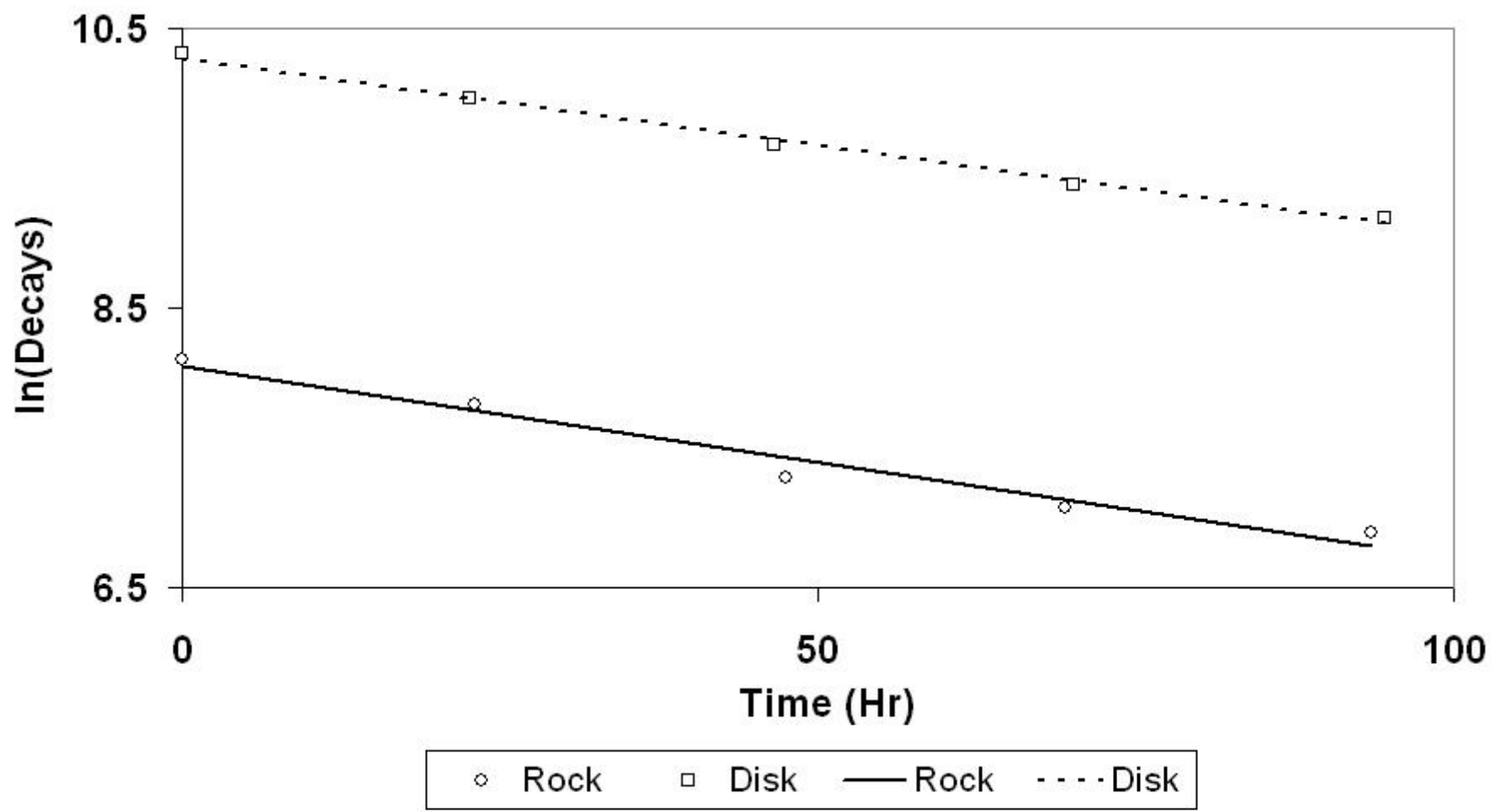